\documentclass[10pt,conference]{IEEEtran}
\usepackage{enumerate}
\usepackage[sort&compress,numbers]{natbib}
\usepackage{graphicx}
\usepackage{subfigure}
\usepackage{epstopdf}
\usepackage{amsthm}
\usepackage{amssymb}
\usepackage{amsmath}
\usepackage[letterpaper, left=0.625in, right=0.625in, bottom=1in, top=0.7in]{geometry}
\usepackage{cleveref}
\usepackage{multirow}
\begin{document}
\title{On the Performance of Dual RIS-assisted V2I Communication under Nakagami-$m$ Fading}
\author{\IEEEauthorblockN{Mohd Hamza Naim Shaikh, Khaled Rabie$^{\circ}$, Xingwang Li$^{\#}$, Theodoros Tsiftsis$^{\dagger}$, and Galymzhan Nauryzbayev} 
\IEEEauthorblockA{School of Engineering and Digital Sciences, Nazarbayev University, Nur-Sultan City, 010000, Kazakhstan\\
$^{\circ}$Department of Engineering, Manchester Metropolitan University, Manchester, M15 6BH, UK\\ 
$^{\#}$School of Physics and Electronic Information Engineering, Henan Polytechnic University, Jiaozuo 454000, China\\
$^{\dagger}$Department of Informatics \& Telecommunications, University of Thessaly, Greece;\\
$^{\dagger}$School of Intelligent Systems Science and Engineering, Jinan University, China\\
Email: \{hamza.shaikh, galymzhan.nauryzbayev\}@nu.edu.kz, $^{\circ}$k.rabie@mmu.ac.uk, \\
$^{\#}$lixingwang@hpu.edu.cn, $^{\dagger}$tsiftsis@ieee.org
}} 
\vspace{-0.5 cm}
\maketitle

\begin{abstract}
Vehicle-to-everything (V2X) connectivity in $5$G-and-beyond communication networks supports the futuristic intelligent transportation system (ITS) by allowing vehicles to intelligently connect with everything. The advent of reconfigurable intelligent surfaces (RISs) has led to realizing the true potential of V2X communication. In this work, we propose a dual RIS-based vehicle-to-infrastructure (V2I) communication scheme. Following that, the performance of the proposed communication scheme is evaluated in terms of deriving the closed-form expressions for outage probability, spectral efficiency and energy efficiency. Finally, the analytical findings are corroborated with simulations which illustrate the superiority of the RIS-assisted vehicular networks.

\end{abstract}

\section{Introduction}
As a key enabler for intelligent transportation systems (ITSs), vehicle-to-everything (V2X) communication has sparked a renewed interest in the research community. V2X encompasses a wide range of wireless technologies such as vehicle-to-pedestrian (V2P), vehicle-to-infrastructure (V2I), and vehicle-to-vehicle (V2V). Additionally, it also includes the vehicular communications with vulnerable road users (VRUs), grid (V2G), network (V2N) and cloud (V2C) \cite{9614348}. The V2X communications will be a critical component of the futuristic connected and self-driving cars, envisioned and enabled by the sixth-generation ($6$G) wireless technologies. Furthermore, the V2X communications will also enhance and transform the quality-of-service (QoS) in terms of  unparalleled user experience, ultra-high road safety and air quality improvement. In addition, a slew of advanced applications will also be supported like platooning, trajectory alignments, exchanging sensor data and high precision maps, and so on \cite{cao2022toward}. Thanks to the enhanced capabilities of $6$G, vehicles will receive accurate safety information, intelligent traffic management support, and innovative infotainment features. Thus, the $6$G services will be used to create a fully automated, autonomous, and ubiquitously connected vehicular network \cite{cheng2020vehicular}.

Recently, reconfigurable intelligent surfaces (RISs) have emerged as a breakthrough technology that offers a great deal in terms of wireless communication \cite{9326394}. Inherently, RIS is a software-defined artificial structure made up of a large number of scattering passive elements, termed as reflecting units (RUs). These RUs are capable to adjust the electromagnetic (EM) properties of a reflected wave that is incident on them. Thus, RIS can use not only this ability to boost the received signal's power, but also the capability to create an additional reflective link to mitigate the impact of blockages. With the large number of RUs, RISs are particularly known to have large spectral and energy efficiency \cite{8741198}. As a result, RIS may be used to improve the quality of vehicular communication through establishing a low-cost, highly energy efficient indirect line-of-sight (LoS) communications \cite{javed2022reliable}.

In \cite{9322158}, the authors investigated the outage performance for RIS-assisted vehicular communication networks. Likewise, the secrecy outage performance of RIS-aided vehicular communications has been studied in \cite{9453160}. RISs were also investigated for detecting VRUs such as cyclists, pedestrians and wheelchair users \cite{9575354}. Specifically, the authors utilized RISs for enhancing the radar visibility for VRUs. Further, in \cite{9144463}, the authors provided a optimization framework for resource allocation in the RIS-aided vehicular communications. Specifically, they jointly optimized the power allocation, RIS reflection coefficients and spectrum allocation for different QoS requirements of the V2V and V2I communication links. Likewise, in \cite{9677910}, the authors discussed a system model where RSU leverages RIS to connect the dark zones, i.e., areas blocked due by obstacles. 
Moreover, a comprehensive overview on the recent advances in $6$G vehicular networks was provided in \cite{rahim, 9563122}, where the authors also described various open challenges and possible research directions.

Motivated by the above, in this work, we investigate the performance of a dual RIS-assisted V2I communication network scenario. Specifically, the proposed scenario considers the uplink transmission where the vehicle is communicating with the base station. To enhance the communication capabilities, the vehicle is supported through two RISs which create a virtual line-of-sight (LoS) link, which, otherwise, was inherently non-LoS (NLoS). The major contributions 
%of this work 
can be summarized as 
\begin{itemize}
    \item Explicitly, we invoked the central limit theorem (CLT) to characterize the received signal-to-noise ratio (SNR) for the proposed dual RIS case. Further, based on this, we derived the closed-form expression for outage probability.
    \item Further, we derived the closed-form expressions for the upper and lower bounds of SE and EE of the proposed dual RIS-assisted V2I communication scenario. 
    \item Finally, as a performance benchmark, the proposed scenario is compared with the single RIS-assisted V2I communication and with RIS V2I communication. The results show the superiority of the proposed scenario of dual RIS-assisted V2I over the single RIS-assisted V2I communication case.  
    \end{itemize}

\section{System Model}

\begin{figure}[!t]
\centering
\includegraphics[width= 7 cm, height = 5 cm]{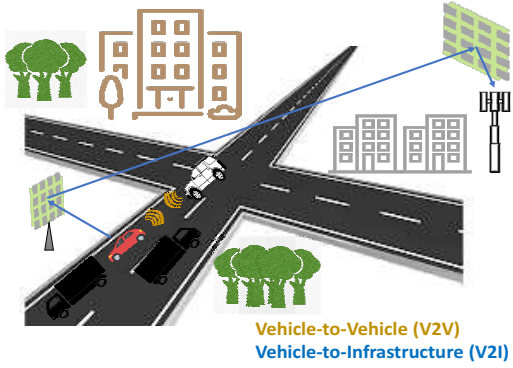}
\caption{Schematic for the considered dual RIS-aided V2I communication.}
\label{fig:Blk1}
 \vspace{-0.5cm}
\end{figure}

As illustrated in Fig. \ref{fig:Blk1}, in this work, we consider a V2I communication model, wherein the vehicular user (V) tries to communicate with a nearby base station (BS). Apart from the direct cellular link, a reflected path through RISs is considered to support this uplink transmission. 
In particular, we consider a dual RIS-assisted uplink V2I transmission with two RISs, one each placed near V and BS both, respectively. 
% end, respectively. 
For the two RISs, the number of RUs is assumed to be $M_1$ and $M_2$ for RIS-1 and RIS-2, respectively, while keeping the total number of RUs unchanged, i.e., $M_1+M_2 = N$, where $N$ is the number of RUs in large RIS for the single RIS scenario, which is the benchmark for comparison. Thus, based on RIS, the following scenarios are considered in this work
\begin{itemize}
        \item \textbf{Dual RIS-assisted Transmission (DRAT)}: In DRAT, the transmission takes place only through the two RISs and the reflected link, as shown in Fig. \ref{fig:Blk1}. 
        \item \textbf{Single RIS-assisted Transmission (SRAT)}: In SRAT, the transmission takes place through single large RIS which is placed near to BS.  
        \item \textbf{Direct Cellular Transmission (DCT)}: In DCT, V communicates with BS directly without utilizing RISs. Thus, the transmission is inherently NLoS and experiences a higher pathloss exponent. This would also serve as the baseline scheme for the performance comparison of the above two cases. 
\end{itemize}

% \subsection{Dual RIS-aided V2I communication scenario}
% As depicted in Fig. \ref{fig:Blk1}, the scenario comprises a typical dense urban environment with a typical link length of around $100$ m. For this scenario, the direct path between V and BS may be blocked, and, to support the transmission, the signal is reflected via two separate RISs, each located near V and BS. This proposed scenario can alleviate the limitation, particularly, in the mmWave and THz frequency bands. For instance, to support high data rate transmission in an V2I or I2V environment over a considerable distance and high frequency, the signal may be attenuated significantly or deeply faded due to multiple obstacles between the propagating paths. 

\subsection{Channel Model}
The channels between V-to-RIS-1 and RIS-2-to-BS can be modeled as deterministic LOS channels as the distances are small and the probability of having LoS is very high. However, the distance between RIS-1 and RIS-2 is large and thus the small scale fading for the channel between the $i$th element of RIS-1 and the $j$th element of RIS-2, denoted by $h^{RR}_{ij}$, is modeled through Nakagami-$m$ fading. Hence, for $i = \{1, 2, \ldots, M_1\}$ and $j = \{1, 2, \ldots, M_2\}$. 
Further, the distances related to the V-to-RIS-1, RIS-1-to-RIS-2 and RIS-2-to-BS links are represented by $d_{1}$, $d_{RR}$ and $d_{2}$, respectively.

\subsection{Received Signal Model} 
The received base-band signal at BS, denoted by $r$, for the dual RIS-aided transmission case can be expressed as 
\begin{align} 
r= \sqrt {\mathcal{B}\,P_{t}} \,\left ({\sum \nolimits _{i=1}^{M_1} \sum \nolimits _{j=1}^{M_2}{e^{j\phi _{i}^{(1) }}h^{R_{}R_{}}_{ij} e^{j\phi _{j}^{(2) }}} }\right)s + N_o, 
\label{eqn_rx_1}
\end{align}
where $P_t$ is the transmit power constraint at V, $\mathcal{B}$ is the distance-dependent pathloss, $s \sim \mathcal{CN}\left(0, 1\right)$ is the transmitted symbol, and $N_o \sim \mathcal{CN}\left(0, \sigma^2\right)$ is the additive white Gaussian noise (AWGN). Further, $\phi_{1}$ and $\phi_{2}$ are the phase of the V-to-RIS1 and RIS2-to-BS channels.
Further, for a link distance $d$, $\mathcal{B}$ at the carrier frequency of $3$ GHz can be given by \cite{8888223}
\begin{align} 
\mathcal{B} (d) \;\mathrm {[dB]}= 
\begin{cases} 
-37.5 - 22 \log_{10}(d/1\,{\mathrm {m}}) & {\mathrm {if~ LOS}}, \\ 
-35.1 - 36.7 \log_{10}(d/1\,{\mathrm {m}}) & {\mathrm {if~ NLOS}}. 
\end{cases} 
\end{align}

Likewise, instantaneous SNR at BS can be formulated as
\begin{equation} 
\gamma =\frac {{{\left |{ \sum \nolimits _{i=1}^{M_1} \sum \nolimits _{j=1}^{M_2}{ \delta _{ij} e^{j\left(\phi _{i}^{(1) }+\phi _{j}^{(2) }-\varphi _{ij}\right)} } }\right |}^{2}}{\mathcal{B}\,P_{t}}}{\sigma^2},
\label{eqn_snr_1}
\end{equation}
where $\delta _{ij}$ and $\varphi _{ij}$ denote the amplitude and phase of $h^{R_{}R_{}}_{ij}$.

\subsubsection{RIS Reflection Parameters}
Now, SNR at BS can be maximized through adjusting the phase at RISs to cancel the resultant phase, i.e., $\phi_{i}^{(1)} + \phi_{j}^{(2)} - \varphi_{ij} = 0$, for $i = \{1, 2, \ldots, M_1\}$ and $j = \{1, 2, \ldots, M_2\}$. Thus, by substituting $\varphi_{ij} = \phi_{i}^{(1)} + \phi_{j}^{(2)}, \forall i,j$, the received signal power at BS can be maximized.  
Consequently, maximized SNR corresponding to the optimal phase can be given as
\begin{align} 
\gamma_{\max } &= \frac {{{\left |{ \sum \nolimits_{i=1}^{M_1} \sum \nolimits_{j=1}^{M_2}{ \delta_{ij} } }\right |}^{2}}{\mathcal{B}\, P_{t}}}{{\sigma^2}} = \frac {A^{2}{\mathcal{B}\,P_{t}}}{\sigma^2} = A^2\, \mathcal{B}\, \Bar{\gamma},
\label{eqn_snr_2}
\end{align}
where 
$A^2 = {\left |{\sum \nolimits _{i=1}^{M_1} \sum \nolimits _{j=1}^{M_2}{\delta_{ij}}}\right |}^{2}$ is the cascaded channel gain provided by RISs, and
$\Bar{\gamma} = P_{t}/\sigma^2$ is transmit SNR.
    %variance due to the .

Likewise, proceeding in the similar way, for the SRAT scenario, maximized SNR at BS can be given as\footnote{For the SRAT scenario, the analysis is similar. Thus, the detailed description is omitted for the sake of brevity. In particular, for SRAT, large RIS with $N$ RUs is present near BS, where $N = M_1+M_2$. Likewise, the RIS-to-BS link is also modeled as Nakagami-$m$ fading with the rest of the parameters being the same, as in DRAT, like transmit power constraint at V, etc.}
\begin{align} 
\hat{\gamma} _{\max } = {\left({ \sum \nolimits _{i=1}^{N}{ \beta_{i} } } \right) }^{2}{\Bar{\gamma}}  = B^2\Bar{\gamma},
\label{eqn_snr_3}
\end{align}
where $\beta_i$ is the amplitude of a channel between RIS and V, denoted by $h^{RU}_i$, i.e., $h_i^{RU} = \beta_i e^{-j\varphi_i}$, and $B^2$ is the corresponding channel gain provided by single RIS. 

\section{Performance Analysis}
This section initially evaluates SNR for the dual RIS-aided V2I scenario. Utilizing the SNR expressions formulated earlier, the outage probability, SE and EE are derived. % are illustrated in the following
%\subsection{Outage Probability}

\subsection{Statistical Characterization of the Dual RIS Channel Gain}
Now utilizing CLT for $M \gg 1$, $A = \sum \nolimits_{i=1}^{M_1} \sum \nolimits_{j=1}^{M_2}{ \delta_{ij} }$ can be approximated through a Gaussian distribution, i.e., $A \sim \mathcal{N}(\mu_y, \sigma_y^2)$  \cite{9148760}, with a probability density function (PDF) given by
\begin{align}
\mathrm{f}_A(y)=\begin{cases}
\frac{1}{\sqrt{2 \pi \sigma_A^2}} \exp{\left(\frac{-\left(y-\mu_A\right)^2}{2\sigma_A^2}\right)}, & \text{ if } y > 0,\\ 
0, & \text{ if } y = 0,
\end{cases}
\label{eqn_pdf_1}
\end{align}
where 
%\begin{align}
    $\mu_A = \sum \nolimits_{i=1}^{M_1} \sum \nolimits _{j=1}^{M_2}{\mu_{ij}}$,
    %\label{mu_1} \\
    $\sigma_A^2 = \sum \nolimits _{i=1}^{M_1} \sum \nolimits _{j=1}^{M_2}{\sigma_{ij}^2}$.
    %\label{sigma_1}
%\end{align}
Here, $\mu_{ij}$ and $\sigma^2_{ij}$ are the mean and variance of the random variable $\delta_{ij}$, which follows the Nakagami-$m$ distribution. Hence, %$\mu_{ij}$ and $\sigma^2_{ij}$ can be defined as 
%\begin{align}
$\mu_{ij} =   \frac {\Gamma(m_{1}+\frac{1}{2})}{\Gamma(m_{1})} \sqrt{\left(\frac{\Omega_{m_{1}}}{m_{1}}\right)}$
%\label{mu_2}
%\end{align}  
and 
%\begin{align}
$\sigma^2_{ij} =  \Omega_{m_1}\left\{1 - \frac{1}{m_{1}}\left(\frac {\Gamma(m_{1}+\frac{1}{2})} {\Gamma(m_{1})}\right)^2\right\}$,
%\label{sigma_2}
%\end{align}
for all $i = \{1, \ldots, M_1\}$ and $j= \{1, \ldots, M_2\}$.    
%where $\mathcal{B}_1$

Likewise the cumulative distribution function (CDF) of $A$ can be derived from its PDF as
\begin{align}
   \mathrm{F}_A(y) \hspace{-0.05cm} = \hspace{-0.05cm} \int_{-\infty}^{y} \mathrm{f}_A(y) \mathrm{d}y \hspace{-0.05cm} = \hspace{-0.05cm}
   \begin{cases} 1 \hspace{-0.05cm} - \hspace{-0.05cm} \mathrm{Q}\left(\frac{y-\mu_A}{\sigma^2_A}\right), & \text{ if } y > 0,\\ 
    0, & \text{ if } y = 0.
    \end{cases}
   \label{eqn_CDF_1}
\end{align}
% The accuracy of this approximation can be verified through the simulation plots, as shown in Fig. \ref{fig:SE5}. % \cite{9095301}. %for different values of $M_1$ and $M_2$. 

% \begin{figure}[!t]
% \centering
% \includegraphics[width= 0.8\columnwidth]{plot_Y_pdfff.eps}
% \caption{PDF of $A$ for different $M$, here $M_1 = M_2 = M$.}
% \label{fig:SE5}
%   % \vspace{-0.3cm}
% \end{figure}

\subsection{Outage Probability} 
The normalized instantaneous rate, denoted by $R_{\rm in}$, for the DRAT scenario can be formulated from \eqref{eqn_snr_2} and expressed as
\begin{align}
R_{\rm in} & = \mathrm{log}_{2}\left( 1+ \gamma_{max} \right) = \mathrm{log}_{2}\left( 1+ A^2\Bar{\gamma}\right).
\label{rate_in}
\end{align}
%where  
Now, the end-to-end outage from V to BS via RIS, denoted by $P_{\rm out}$, can be defined in terms of a rate threshold, $R_{\rm th}$, as   
%\begin{equation}
%P_{\rm out}=\mathrm{Pr} \left[ R_{\rm in}< R_{\rm th} \right].
%\label{eqn_out_1}
%\end{equation}
%From \eqref{rate_in}, $P_{\rm out}$ can be given as
\begin{align}
P_{\rm out}&= \mathrm{Pr} \left[ R_{\rm in}< R_{\rm th} \right] = \mathrm{Pr} \left[\mathrm{log}_{2}\left(1+A^2\Bar{\gamma}\right) < R_{\rm th} \right] \nonumber  \\ 
&=\mathrm{Pr} \left [A < \sqrt{\frac{ 2^{R_{\rm th}}-1}{\Bar{\gamma}}} \right ] 
=\mathrm{Pr} \left [A < \Upsilon_{\rm th} \right ],
\label{eqn_out_2}
\end{align}
where $\Upsilon_{\rm th}=\sqrt{\frac{2^{R_{\rm th}}-1}{\Bar{\gamma}}}$. Thus, the closed-form expression of the outage probability DRAT can be evaluated as 
\begin{align}
P_{\rm out} =& \int_{0}^{\Upsilon_{\rm th}} \mathrm{f}_A(y)\mathrm{d}y, \nonumber \\
 =&  \mathrm{F}_A\left(\Upsilon_{\rm th}\right) = 1 - \mathrm{Q}\left(\frac{\Upsilon_{\rm th}-\mu_A}{\sigma^2_A}\right).  \label{Pout_SR}
\end{align}

\begin{figure*}
\centering
\begin{align}
    SE_l = \log_2 \left[ 1 +  \Bar{\gamma}\, \mathcal{B}\,  \frac{M_1M_2 \, \Omega_{m_1} \left\{1 + {\frac{\left(M_1\,M_2 - 1\right)}{m_1}}\; \left( \frac {\Gamma(m_{1}+\frac{1}{2})} {\Gamma(m_{1})} \right)^2 \right\}^3} {2\left\{1+ {\frac{\left(2M_1\,M_2 - 1\right)}{m_1}}\; \left( \frac {\Gamma(m_{1}+\frac{1}{2})} {\Gamma(m_{1})} \right)^2 \right\}\left\{1- \frac{1}{m_1} \left(\frac{\Gamma(m_{1}+\frac{1}{2})} {\Gamma(m_{1})} \right)^2\right\} \hspace{-0.05cm} + \hspace{-0.05cm} \left\{1 + \frac{ \left(M_1\,M_2 - 1\right)}{m_1} \left(\frac{\Gamma(m_{1}+\frac{1}{2})} {\Gamma(m_{1})} \right)^2 \right\}^2}\right]  \tag{16}
    \label{24}
\end{align}
\hrule
\vspace{-0.3cm}
\end{figure*}

\subsection{Spectral Efficiency}
SE for the DRAT scenario can be defined from \eqref{rate_in} as
\begin{align}
SE_{} =& \mathrm{E} \left[R_{\rm in} \right] = \mathrm{E} \left[ \mathrm{log}_{2} \left( 1+ A^2\,\mathcal{B}\, \Bar{\gamma}\right)\right], \nonumber \\
%\label{eqn_SE_1}
%\end{align}
%Now, the closed-form expression for SE of the DRAT scenario can be evaluated by solving the expectation over the distribution of a channel gain $A$. 
%\begin{align}
   =& \int_{0}^{\infty} \mathrm{log}_{2}\left ( 1+ y^2 \,\mathcal{B}\, \Bar{\gamma}\right ) \mathrm{f}_A(y)\mathrm{d}y.
\label{eqn_SE_2}
\end{align}
The exact derivation of the integral in \eqref{eqn_SE_2} is mathematically intractable, and thus a closed-form expression may not be derived. Hence, we resort to approximate SE with tight upper and lower bounds by invoking Jensen’s inequality.

\subsubsection{Upper Bound}
Applying Jensen’s inequality, we define the upper bound for SE as ${SE}_{u}$, where 
%\begin{align} 
$SE \leq {SE}_{u}$. 
%\end{align}
Now, ${SE}_u$ can be evaluated from \eqref{eqn_SE_2} as
\begin{align}
{SE}_{u} = \log_2 \left( 1 + \Bar{\gamma} \,\mathcal{B}\, \mathrm{E} \left[ A^2\right] \right),
\label{eqn_SE_3}
\end{align}
and expressed as
\begin{multline}
    SE_u = \log_2 \left[ 1 +  \Bar{\gamma}\,\mathcal{B}\, M_1M_2\,\Omega_{m_1}\right. \\
    \left. \times \left\{1 +  {\frac{\left(M_1\,M_2 - 1\right)}{m_1}}\; \left( \frac {\Gamma(m_{1}+\frac{1}{2})} {\Gamma(m_{1})} \right)^2 \right\}\right].
    \label{22}
\end{multline}

%be as shown in \eqref{22}, at the top of this page.
%\setcounter{equation}{22}

\textit{Evaluation of Upper Bound}: In \eqref{eqn_SE_3}, $\mathrm{E} \left[ A^2 \right]$ can be evaluated utilizing the relation $\mathrm{Var}\left[X\right] = \mathrm{E} \left[ X^2\right] - \left(\mathrm{E} \left[ X\right]\right)^2$ as
\begin{align}
\mathrm{E} \left[ A^2\right]  =&  \mathrm{Var} \left[ A\right] + \left( \mathrm{E} \left[ A\right]\right)^2 = \sigma^2_A + \mu_A^2.
\end{align}
After substituting the values of $\mu_A^2$ and $\sigma^2_A$ in \eqref{eqn_SE_3}, the upper bound for DRAT-based SE can be evaluated.

\subsubsection{Lower Bound}
Likewise, we define the lower bound for SE as ${SE}_{l}$, where 
%\begin{align} 
$SE \geq {SE}_{l}$. 
%\end{align}
Now, ${SE}_l$ can again be be defined from \eqref{eqn_SE_2} as
\begin{align}
{SE}_{l} = \log_2 \left( 1 + \frac{\Bar{\gamma} \,\mathcal{B}}{\mathrm{E} \left[ \frac{1}{A^2}\right]} \right),
\label{eqn_SE_5}
\end{align}
and expressed as given in (16), on the top of next page. 
\setcounter{equation}{16}

\textit{Evaluation of Lower Bound}: In \eqref{eqn_SE_5}, the expectation $\mathrm{E} \left[ {1}/{A^2} \right]$ can be solved utilizing the Taylor series expansion and approximated as \cite{9148760}
\begin{align}
    \mathrm{E}\left[ \frac{1}{A^2}\right] \approx \frac{1}{\mathrm{E}\left[{A^2}\right]} + \frac{\mathrm{Var}\left[{A^2}\right]}{\left[\mathrm{E}\left[{A^2}\right]\right]^3}. 
\end{align}
Since the statistical characteristics of $A$ is known to be Gaussian distributed (as discussed earlier in subsection A), $A^2$ will follow a non-central chi-square distribution. Thus, the mean and variance of $A^2$ can be expressed as
\begin{align}
\mathrm{Var}\left[A^2\right] &= 2\,\sigma^2_A\,\left(\sigma^2_A+2\,\mu^2_A\right), \\
\mathrm{E}\left[A^2\right] &= \sigma^2_A+\mu^2_A,
\end{align}
respectively. Thus, utilizing these expressions and substituting the values of $\mu_A^2$ and $\sigma^2_A$, the lower bound for SE of the DRAT scenario can be evaluated.

\subsubsection{Approximation for Large $M$} 
We define $\overline{SE}$ as approximate SE (ASE) for large $M_1$ and $M_2$. Now, with the upper and lower bounds of SE of the DRAT scenario, exact SE lies in-between and can be expressed as
\begin{align}
SE_l\leq SE \leq SE_u.
\label{SE_bound_1}
\end{align}
However, for larger $M_1$ and $M_2$, i.e., $M_1, M_2 \gg 1$, both $SE_l$ and $SE_u$ converge to $\overline{SE}$. Thus, ASE can be given as 
\begin{align}
    \overline{SE} = \log_2 \left[ 1 +  \Bar{\gamma} \,\mathcal{B}\,  {\frac{M_1^2\,M_2^2\,\Omega_{m_1}}{m_1}}\; \left( \frac {\Gamma(m_{1}+\frac{1}{2})} {\Gamma(m_{1})} \right)^2 \right].
    \label{SE_bound_2}
\end{align}

It can be noted from \eqref{SE_bound_2} that, through utilizing dual RIS, the fourth order channel gain can be realized, i.e., $M_1^2\,M_2^2$, whereas, for single RIS, the maximum channel gain is of the second order, i.e., $N^2$. %This shows the potential of dual RIS  

\subsection{Energy Efficiency} 
Now, EE of the dual RIS-aided system is defined as the ratio of SE over the total power consumed and can be expressed as % \cite{9298754}, , $P_{tot}$. 
%\begin{equation} 
$EE = \frac{SE}{P_{\rm tot}}$,
%\label{eqn_EE_1}
%\end{equation}
where $P_{\rm tot}$ denotes the total power consumed, which consists of the transmit power, the circuit power consumption at BS and V, and the power consumed at RIS. 
Considering all the power consumed, the EE in can be expressed 
\begin{align}
EE = \frac{SE}{(1+ \xi) P_{t} + P_{V}^c + (M_1+M_2)P^c_{RIS} + P_{BS}^c},
\label{eqn_EE_2}
\end{align}
where $P_{RIS}^{c}$ denotes the power utilized by each RU, $\xi = \frac{1}{\omega}$ and $\omega$ is the drain efficiency of HPA. Likewise, $P_{V}^c$, i.e., the power consumed in other circuit components excluding HPA at V and $P_{BS}^c$ is the circuit power consumption at BS.

This completes the analytical derivation of the outage, SE, and EE for DRAT of the uplink of V2I communication.

\begin{table}[!t]
\renewcommand{\arraystretch}{1.3}
\centering
\caption{Simulation Parameters}
\label{t1}
\begin{tabular}{||c | c||}
\hline
\textbf{Parameter} & \textbf{Simulation Values} \\ [0.5ex]\hline \hline
 Circuit Power & $P_{BS}$=10 dBm, $P_U$=10 dBm \cite{8741198} 
 \\ \hline
Fading Parameter for DRAT    & $m_{1}$= 10  \\ \hline
Fading Parameter for Direct Links     & $m_{3}$= 1  \\ \hline
RIS Power Consumption & $P_{RE}$  = 10 dBm  \cite{8741198} 
\\  \hline
HPA Power Consumption Factor & $\alpha$ = 1.2 %\cite{9298754}
\\ \hline
Noise Floor & $\sigma^2$ = -120 dBm \\ 
[1ex]\hline
\end{tabular}
\label{Tab1}
\vspace{-0.4cm}
\end{table}

% \begin{table}[!t]
% \renewcommand{\arraystretch}{1.3}
%  \centering
% \caption{Evaluation Matrix}
% \label{tabb}
% \begin{tabular}{||l|l|l|l|l||} \hline
% % \multirow{2}{*}{Simulation Parameter} & \multicolumn{4}{l||} {M=5}  \\   \cline{2-5}
% \textbf{Parameter} & $SE_l$ & $SE$ & $SE_u$  & $\overline{SE}$   \\ \hline \hline
% $\Bar{\gamma}$ = 30 dB, $M=10$ & 1.5027 & 1.5037 & 1.5038 & 1.5034\\
% $\Bar{\gamma}$ = 15 dB, $M=20$ & 0.9474 & 0.9474 & 0.9476 & 0.9475 \\
% $\Bar{\gamma}$ = 30 dB, $M=20$ & 4.9240 & 4.9242 & 4.9243 & 4.9242\\
% $\Bar{\gamma}$ = 15 dB, $M=50$ & 5.2199 & 5.2201 & 5.2200 & 5.2200\\
% $\Bar{\gamma}$ = 30 dB, $M=50$ & 10.1648 & 10.1649 & 10.1649 & 10.1649\\              \hline      
% \end{tabular}
% \vspace{-0.4cm}
% \end{table}

\section{Simulation Results}
This section discusses and presents the simulation results for the performance of the dual RIS-assisted V2I communication. %For the direct links between RISs, the Nakagami fading parameter is assumed to be $m_{1}  = 10$ for DRAT. Likewise, for the V-to-RIS reflected links in the SRAT scenario $m_{2} = 5$. And for the DCT scenario the value of Nakagami fading parameter is assumed to be $m_{3} = 10$ for V-to-I NLoS link, if not specified otherwise. 
Further, the results for the SRAT and DCT scenarios are presented for the sake of comparison.
The distances between V-to-RIS1, RIS1-to-RIS2 and RIS2-to-BS are assumed to be $5$, $100$ and $5$ meters, respectively. Similarly for the simulation purpose, $M = M_1 = M_2$ and $N$ is taken as to be $N= 2\,M$, in order to maintain the fairness in the comparison. The rest of the simulation parameters are summarized in Table \ref{Tab1}.

\begin{figure}[!t]
\centering
\includegraphics[width= 7 cm, height = 4 cm]{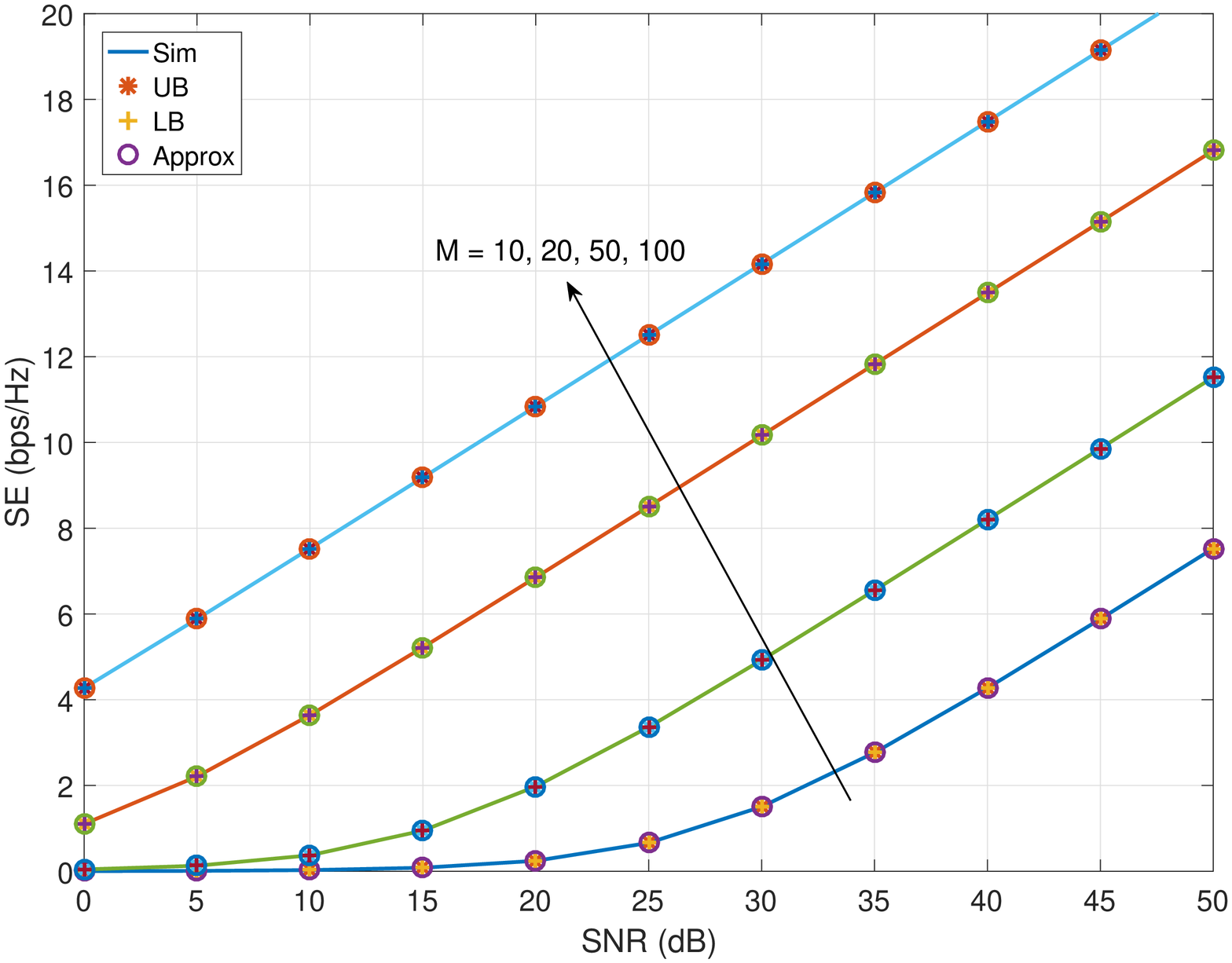}
\caption{SE with respect to $\Bar{\gamma}$ for different $M$ of the proposed DRAT scenario.}
\label{fig_SE1}
\vspace{-0.4cm}
\end{figure}

\begin{figure}[!t]
\centering
\includegraphics[width= 7 cm, height = 4 cm]{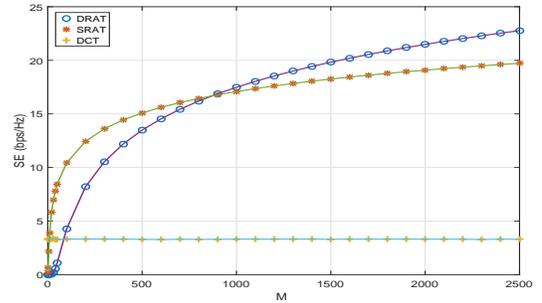}
\caption{SE with respect to $M$ for the proposed DRAT scenario.}
\label{fig_SE2}
\vspace{-0.4cm}
\end{figure}

\begin{figure}[!t]
\centering
\includegraphics[width= 7 cm, height = 4 cm]{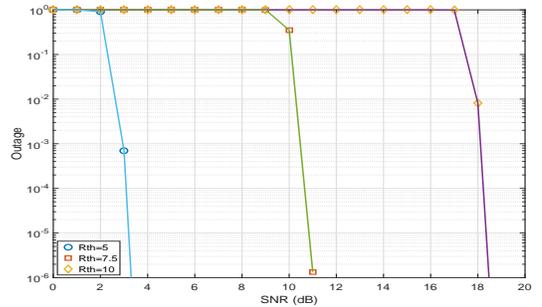}
\caption{Outage with respect to $P_t$ for different rate thresholds for DRAT.}
\label{Fig_Out1}
\vspace{-0.4cm}
\end{figure} 

\begin{figure}[!t]
\centering
\subfigure[EE with respect to $M$, here $P_t = 10$ dBm.]{\label{Fig_EE1}
\includegraphics[width= 7 cm, height = 4 cm]{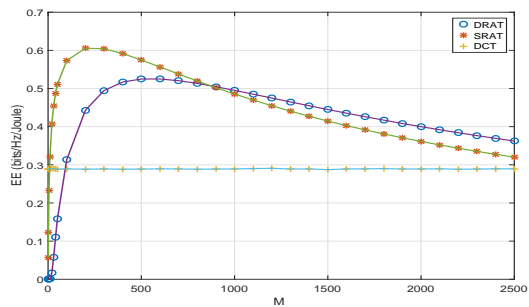}}
\subfigure[EE with respect to SNR, here $M=1000$.]{\label{Fig_EE2}
\includegraphics[width= 7 cm, height = 4 cm]{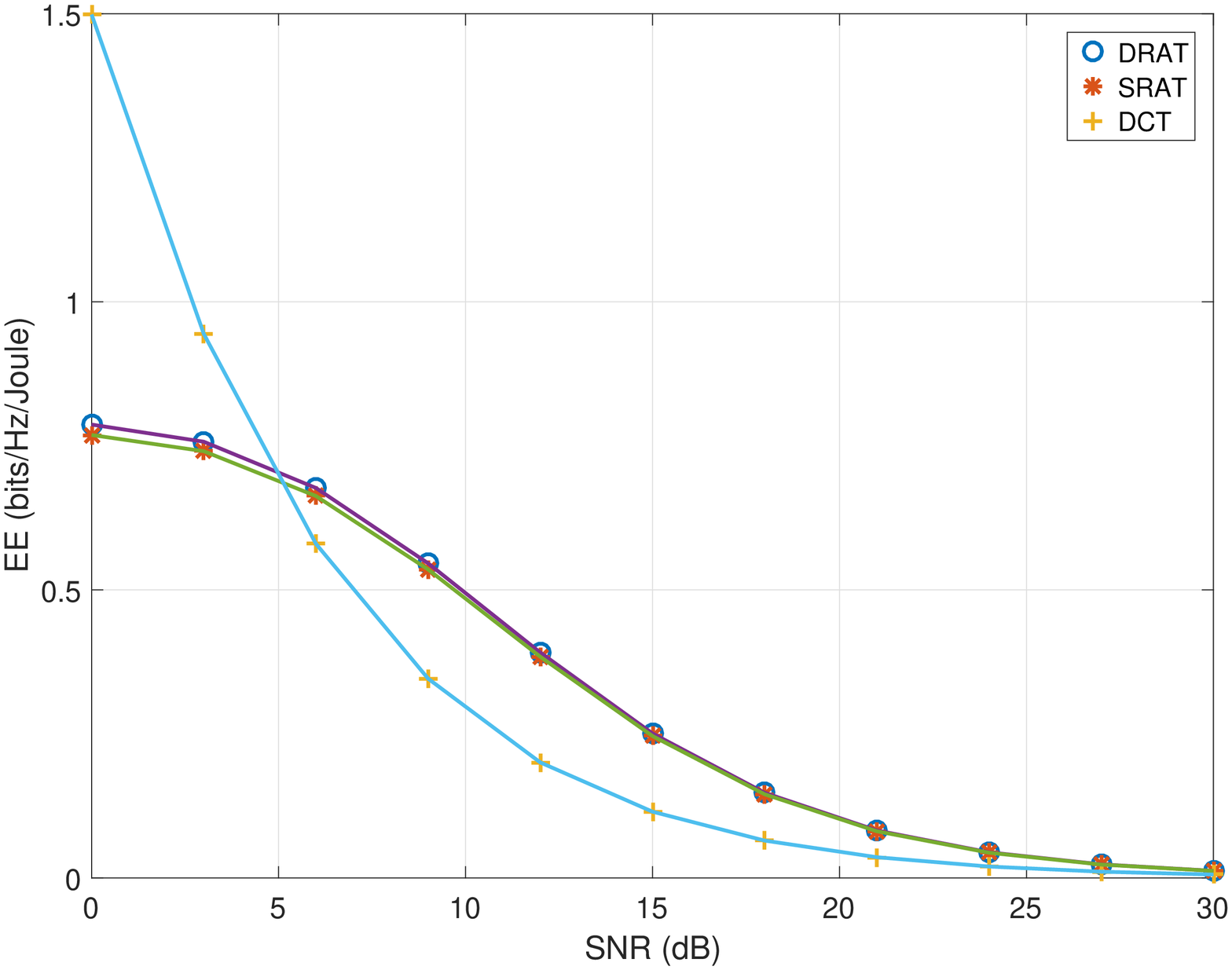}}
\caption{EE comparison for DRAT with respect to the SRAT and DCT scenarios.}
\label{Fig_EEE1}
\vspace{-0.4cm}
\end{figure}

Fig. \ref{fig_SE1} shows the SE performance for the DRAT scenario, where the solid lines without marker points show the exact (simulation) performance of DRAT, whereas the markers show the analytically derived upper and lower bounds on SE. Additionally, ASE for large $M$ is also plotted. The simulation verifies that the derived upper and lower bounds are quite tight as the analytically derived bounds are remarkably close to the actual performance. 
%Additionally, for the ease of purview, Table \ref{tabb} summarizes SE and its bounds for the DRAT scenario. 
Further, it can also be noted that the difference between exact SE and ASE (as shown in (21)) diminishes as $M$ increases. 
For instance, at $M=10$ and $\Bar{\gamma} = 30$ dB, $SE$ is $1.5037$ bps/Hz whereas $\overline{SE}$ is $1.5034$ bps/Hz; however, at $M =50$ and $\Bar{\gamma}=15$ dB, $SE$ is $5.2201$ whereas $\overline{SE}$ is $5.2200$ bps/Hz. Thus, it shows that the bounds are quite accurate and near to the exact simulation value.  

Fig. \ref{fig_SE2} shows the SE results for the DRAT scenario, and compares them with the SRAT and DCT scenarios. Specifically, it shows SE for a varying number of RUs. The following observations can be easily inferred from this plot: 1) Apart from smaller $M$, SE of DRAT is always better than SE of the SRAT scheme due to the fourth order gain provided by dual RIS. This can also be inferred from the analytical evaluation in (31). 2) Due to the multiplicative pathloss, for less number of RUs, i.e., smaller $M$, the DCT scenario may provide better SE performance than the RIS-reflected link for both DRAT and SRAT scenarios. However, as the number of RUs increases, the RIS-based scenarios outperform DCT. 3) Similar to single RIS, dual RIS-based DRAT also suffers from the multiplicative effect of pathloss. Thus, for smaller RUs, the SRAT scenario shows better SE than the DRAT one. 4) As the number of RUs increases sufficiently, DRAT outperforms SRAT significantly.

Fig. \ref{Fig_Out1} shows the outage probability of the DRAT scenario for three different rate thresholds, i.e., $R_{\rm th} = \{5, 7.5, 10\}$ bps/Hz. As evident from the result, the outage can be improved either through increasing the transmit power or the number of RUs. Since, the transmit power at BS is usually constrained, RIS provides an alternate to improve the outage through increasing RUs, instead of increasing the transmit power. Thus, to circumvent the power constraint, the number of RUs at RIS can be scaled accordingly.

Fig. \ref{Fig_EEE1} shows the EE results of the DRAT scenario, the EE plots of the SRAT and DCT scenarios are also plotted here for comparison. Specifically, in Fig. \ref{Fig_EE1}, the performance is with respect to $M$, while in Fig. \ref{Fig_EE2}, the EE curve is plotted against SNR. It can be observed that, for large $M$, the DRAT scenario is the most energy-efficient. Although, for smaller $M$, single RIS provides better EE; this is due to the fact that the received signal of the dual RIS-reflected link suffers from the multiplicative pathloss that can be mitigated by by large $M$.  

From the above results on SE and EE, it can be easily inferred that the proposed DRAT scheme outperforms SRAT in terms of both SE as well as EE. Similarly, the above results also show that, for a fixed rate requirement, DRAT requires lower transmit power and hence is more energy efficient.

\section{Conclusion}
V2X has opened up a slew of novel possibilities in the wireless vehicular communication arena, but its potential for enabling true ITS has yet to be explored completely, despite its significant importance in the safety of autonomous driving.
In this work, we have envisioned the integration of RIS into vehicular networks to realize the true potential in enhancing the performance of the V2I communication. Specifically, we have evaluated the performance of a dual-RIS assisted V2I uplink communication scenario in terms of the outage probability, SE and EE. Novel closed-form expressions are derived and verified through the extensive numerical simulations. The results show a significant gain in the performance can be achieved through the proposed RIS scenario. 

\section{Acknowledgement}
This work was supported by the Nazarbayev University CRP Grant no. 11022021CRP1513.

\bibliographystyle{IEEEtran}
{\footnotesize
\bibliography{Bibil}}

% Generated by IEEEtran.bst, version: 1.14 (2015/08/26)
\begin{thebibliography}{10}
\providecommand{\url}[1]{#1}
\csname url@samestyle\endcsname
\providecommand{\newblock}{\relax}
\providecommand{\bibinfo}[2]{#2}
\providecommand{\BIBentrySTDinterwordspacing}{\spaceskip=0pt\relax}
\providecommand{\BIBentryALTinterwordstretchfactor}{4}
\providecommand{\BIBentryALTinterwordspacing}{\spaceskip=\fontdimen2\font plus
\BIBentryALTinterwordstretchfactor\fontdimen3\font minus
  \fontdimen4\font\relax}
\providecommand{\BIBforeignlanguage}[2]{{%
\expandafter\ifx\csname l@#1\endcsname\relax
\typeout{** WARNING: IEEEtran.bst: No hyphenation pattern has been}%
\typeout{** loaded for the language `#1'. Using the pattern for}%
\typeout{** the default language instead.}%
\else
\language=\csname l@#1\endcsname
\fi
#2}}
\providecommand{\BIBdecl}{\relax}
\BIBdecl

\bibitem{9614348}
J.~Wang, K.~Zhu, and E.~Hossain, ``Green internet of vehicles {(IoV)} in the
  {6G} era: Toward sustainable vehicular communications and networking,''
  \emph{IEEE Trans. Green Commun. Netw.}, vol.~6, no.~1, pp. 391--423, Mar.
  2022.

\bibitem{cao2022toward}
Y.~Cao, S.~Xu, J.~Liu, and N.~Kato, ``Toward smart and secure {V2X}
  communication in {5G} and beyond: A {UAV}-enabled aerial intelligent
  reflecting surface solution,'' \emph{IEEE Veh. Tech. Mag.}, 2022.

\bibitem{cheng2020vehicular}
X.~Cheng, Z.~Huang, and S.~Chen, ``Vehicular communication channel measurement,
  modelling, and application for beyond {5G} and {6G},'' \emph{IET Commun.},
  vol.~14, no.~19, pp. 3303--3311, 2020.

\bibitem{9326394}
Q.~Wu, S.~Zhang, B.~Zheng, C.~You, and R.~Zhang, ``Intelligent reflecting
  surface-aided wireless communications: A tutorial,'' \emph{IEEE Trans.
  Commun.}, vol.~69, no.~5, pp. 3313--3351, May 2021.

\bibitem{8741198}
C.~{Huang}, A.~{Zappone}, G.~C. {Alexandropoulos}, M.~{Debbah}, and C.~{Yuen},
  ``Reconfigurable intelligent surfaces for energy efficiency in wireless
  communication,'' \emph{IEEE Trans. Wireless Commun.}, vol.~18, no.~8, pp.
  4157--4170, Aug. 2019.

\bibitem{javed2022reliable}
M.~A. Javed \emph{et~al.}, ``Reliable communications for cybertwin driven {6G
  IoVs} using intelligent reflecting surfaces,'' \emph{IEEE Trans. Ind. Info.},
  2022.

\bibitem{9322158}
J.~Wang \emph{et~al.}, ``Outage analysis for intelligent reflecting surface
  assisted vehicular communication networks,'' in \emph{IEEE Global Commun.
  Conf.}, 2020, pp. 1--6.

\bibitem{9453160}
Y.~Ai \emph{et~al.}, ``Secure vehicular communications through reconfigurable
  intelligent surfaces,'' \emph{IEEE Trans. Veh. Tech.}, vol.~70, no.~7, pp.
  7272--7276, Jul. 2021.

\bibitem{9575354}
S.~K. Dehkordi and G.~Caire, ``Reconfigurable propagation environment for
  enhancing vulnerable road users' visibility to automotive radar,'' in
  \emph{IEEE Intelligent Veh. Symp. (IV)}, 2021, pp. 1523--1528.

\bibitem{9144463}
Y.~Chen \emph{et~al.}, ``Resource allocation for intelligent reflecting surface
  aided vehicular communications,'' \emph{IEEE Trans. Veh. Tech.}, vol.~69,
  no.~10, pp. 12\,321--12\,326, Oct. 2020.

\bibitem{9677910}
A.~Al-Hilo \emph{et~al.}, ``Reconfigurable intelligent surface enabled
  vehicular communication: Joint user scheduling and passive beamforming,''
  \emph{IEEE Trans. Veh. Tech.}, vol.~71, no.~3, pp. 2333--2345, Mar. 2022.

\bibitem{rahim}
\BIBentryALTinterwordspacing
M.~Noor-A-Rahim \emph{et~al.}, ``{6G} for vehicle-to-everything {(V2X)}
  communications: Enabling technologies, challenges, and opportunities,'' 2020.
  [Online]. Available: \url{https://arxiv.org/abs/2012.07753}
\BIBentrySTDinterwordspacing

\bibitem{9563122}
Y.~Zhu \emph{et~al.}, ``Intelligent reflecting surface-aided vehicular networks
  toward {6G}: Vision, proposal, and future directions,'' \emph{IEEE Veh. Tech.
  Mag.}, vol.~16, no.~4, pp. 48--56, Dec. 2021.

\bibitem{8888223}
E.~{Bjornson}, O.~{Ozdogan}, and E.~G. {Larsson}, ``Intelligent reflecting
  surface versus decode-and-forward: How large surfaces are needed to beat
  relaying?'' \emph{IEEE Wireless Commun. Lett.}, vol.~9, no.~2, pp. 244--248,
  2020.

\bibitem{9148760}
D.~Kudathanthirige, D.~Gunasinghe, and G.~Amarasuriya, ``Performance analysis
  of intelligent reflective surfaces for wireless communication,'' in
  \emph{IEEE Int. Conf. Commun. (ICC)}, 2020, pp. 1--6.

\end{thebibliography}
\end{document}